\documentclass[preprints,article,accept,moreauthors,pdftex]{Definitions/mdpi}

\usepackage{tikz}
\usetikzlibrary{shadows,arrows}

\pgfdeclarelayer{background}
\pgfdeclarelayer{foreground}
\pgfsetlayers{background,main,foreground}
 
\tikzstyle{materia}=[draw, fill=blue!20, text width=6.0em, text centered, minimum height=1.5em,drop shadow]
\tikzstyle{Block} = [materia, text width=6em, minimum width=10em,
  minimum height=3em, rounded corners, drop shadow, text centered]
\tikzstyle{DimBlock} = [materia, text width=10em, minimum width=10em,
  minimum height=3em, rounded corners, drop shadow, text centered]
\tikzstyle{texto} = [above, text width=6em, text centered]
\tikzstyle{line} = [draw, thick, color=black!80, -latex']

\newcommand{\inputblock}[2]{node (inp#1) [Block]
  {Input {\scriptsize\textit{#2}}}}
\newcommand{\bldimension}[2]{node (bld#1) [DimBlock]
  {{} {\textbf{#2}}}}
\newcommand{\block}[2]{node (p#1) [Block]
  {{} {{#2}}}}
\newcommand{\flattenblock}[2]{node (flb#1) [Block]
  {Flatten {\scriptsize\textit{#2}}}}
\newcommand{\denseblock}[2]{node (dnb#1) [Block]
  {Dense #1 {\scriptsize\textit{#2}}}}

\newcommand{\background}[5]{%
  \begin{pgfonlayer}{background}
    \path (#1.west |- #2.north)+(-0.5,0.5) node (a1) {};
    \path (#3.east |- #4.south)+(+0.5,-0.5) node (a2) {};
    \path[fill=yellow!20,rounded corners, draw=black!50, dashed]
      (a1) rectangle (a2);
    \path (a1.east |- a1.south)+(0.8,-0.3) node (u1)[texto]
      {\scriptsize\textit{#5}};
  \end{pgfonlayer}}
\usepackage{verbatim}
\usepackage{forest}
\usepackage{pgfplots}
\pgfplotsset{compat=1.14}

\pgfplotsset{every axis/.style={scale only axis}}

\firstpage{1} 
\pubvolume{xx}
\issuenum{1}
\articlenumber{5}
\pubyear{2019}
\copyrightyear{2019}
\history{Received: date; Accepted: date; Published: date}

\Title{Exploring the Possibility of a Recovery of Physics Process Properties from a Neural Network Model}

\Author{Marko Jercic $^{1}$*, Nikola Poljak$^{1}$}

\AuthorNames{Marko Jercic, Nikola Poljak}

\address[1]{%
$^{1}$ \quad Department of Physics, Faculty of Science, University of Zagreb}

\corres{Correspondence: mjercic@phy.hr}

\abstract{The application of machine learning methods to particle physics often doesn't provide enough understanding of the underlying physics. An interpretable model which provides a way to improve our knowledge of the mechanism governing a physical system directly from the data can be very useful. In this paper, we introduce a simple artificial physical generator based on the Quantum chromodynamical (QCD) fragmentation process. The data simulated from the generator are then passed to a neural network model which we base only on the partial knowledge of the generator. We aim to see if the interpretation of the generated data can provide the probability distributions of basic processes of such a physical system. This way, some of the information we omitted from the network model on purpose is recovered. We believe this approach can be beneficial in the analysis of real QCD processes.}


\begin{document}

\section{Introduction}
Modern particle physics has the potential to answer many open fundamental questions, such as the unification of forces, the nature of dark matter or the neutrino masses. To answer these, we turn to data collected by particle accelerators, such as the Large Hadron Collider (LHC) at CERN. These data are collected by detectors which register signals coming from a collision of particles such as protons or lead nuclei. They are almost exclusively complex and of high dimensionality, so untangling them requires a certain level of understanding of the underlying processes that produce them.

The traditional analysis techniques employed in the high energy physics community use sequences of decisions to extract relevant information. The determination of the statistical significance of the extracted quantities then determine if the data yield a new result or not. This approach is usually limited to a single variable, such as the invariant mass of the system. When more than one variable is considered, a multivariate approach is used, which is already a form of a machine learning technique. Lately, a larger number of these techniques are being implemented in high energy physics data analyses, typically including boosted decision trees, genetic algorithms, random forests or artificial neural networks.

This approach to analysis should be natural, since the data resulting from a particle interaction are fundamentally probabilistic due to the quantum mechanical nature of particle collisions. In this sense, the classical approach to data analysis poses a problem because the statistical model describing them can not be known explicitly in terms of an equation that can be analytically evaluated. To make matters worse, even though we have a good model describing the particle interactions (namely quantum chromodynamics), it is inherently non-perturbative and we can not calculate what it predicts in a certain collision. So, to interpret the collected data we turn to large samples of simulated data generated by stochastic simulation tools such as PYTHIA \cite{pythia} which try to describe the relevant physics within a nucleus–nucleus collision. However, they have their drawbacks in not being exact, but instead relying heavily on Monte Carlo methods. Even though the knowledge incorporated in the simulators is regularly reinforced with new observations from data, one can never expect the complete physical truth from them.

Considering the fact that we can't rely entirely on simulated data, we wanted to develop an interpretable model that will provide a way to improve our knowledge of the mechanisms governing particle collisions. We introduce a simple artificial jet generator based only on generalized conservation laws. The simulated data are then passed to a neural network model based only on the partial knowledge of the generator. We try to interpret the generated data and obtain the probability distributions of basic processes of such a physical system, thus recovering some of the information we omitted from the network model. To do this, we make use of the Neyman-Pearson lemma \cite{NeyPear}, which is an approach that has been proposed by several authors lately \cite{NNNP1, NNNP2}. The developed method could be extended to real data from the LHC, with the hope of gaining new insight on real QCD processes.

This paper is organized as follows: in the Results section we describe how our data is generated and propose the use of the Neyman-Pearson lemma to obtain the underlying probabilities of the data distributions. To do this, we use a neural network classifier and a ``guess'' dataset. We quantify the differences in the obtained and the original probabilities and present them along with the obtained distributions. In the Discussion section we give a conclusion which follows from these results and present the implications for future research. We conclude the paper with the Materials and Methods section, where we detail the methodology used, should someone want to recreate the results on their own.

\section{Results}

\subsection{The jet generator}
\begin{figure}[h!t!]
\centering
\includegraphics[width=15cm]{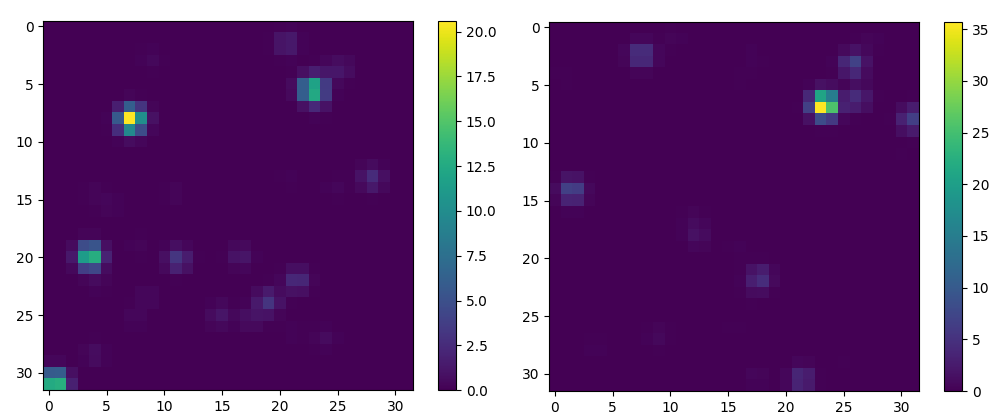}
\caption{Two examples of jet images generated by the procedure outlined in the text. The $x$ and $y$ axes of the graphs correspond to the azimuthal angle $\phi$ and the polar angle $\theta$ with respect to the origin. The full solid angle is mapped on these graphs, with 32 bins used for each angle. The color values in these graphs correspond to the energies of the final particles, with the energy of the original particle set to 100. (\textbf{a}) The left panel shows an image of a jet generated with a probability distribution of gluon momenta radiated by a quark. (\textbf{b}) The right panel shows an image of jet generated with a different probability distribution. Visually, one can note that the left jet has particles whose energies are closer to one another compared to the right jet, in which one particle is of distinctly higher energy.}
\label{JetImage}
\end{figure}

To begin with, we create a sample of data based on a simple physical process which will mimic the data obtained from particle collider experiments. We start with a particle at rest which decays into two particles. The energies and the momenta of these particles are determined by a selected probability distribution, in this case the distribution of gluon momenta radiated by a quark \cite{AltPar}. The spatial distribution of the decay products is uniform in space. After the first decay, the procedure repeats iteratively as described in the Methodology section. The decay procedure stops when either of two conditions is met; if the decay particle mass falls below a preset threshold, or a certain number of decays has been reached. For simplicity, all the decays are considered to happen in the same point in space. The list of final decay particles now forms a $n$-tuple that contains the energies, the momenta and the directions of the $n$ particles. We call this entity a jet. To visualise it, we create a histogram whose axes represent the direction of a particle in space. The histogram axes represent the azimuthal angle $\phi$ and the polar angle $\theta$ of a particle. The color of a pixel in the histogram corresponds to either the energy or the momentum of the particle traveling in that direction in space. An example of the jet generator tree with modified parameters is given in Appendix A. Two examples of jet images are given on figure (\ref{JetImage}).

\subsection{The Neyman–Pearson lemma}

Let us now forget the decay probability distributions implemented in the data we created. We would like to retrieve them by guessing some of their general characteristics. To do this, we use a neural network to differentiate jet images from the created data and jet images from a ``guess'' distribution. The idea is the following: first, a number of jets following a known decay probability distribution $p_{\textrm{real}}$ is created. In our case, this distribution is either the particle energy or the particle momentum distribution, but the arguments we present are valid for any probability distribution. Next, we create another set with the same number of jets in the same manner, this time following a different probability distribution we call $p_{\textrm{guess}}$. 

Assume you are performing a hypothesis test between $H_{0}:p = p_{\textrm{real}}(z)$ and $H_{1}:p=p_{\textrm{guess}}(z)$ using a likelihood-ratio test. The Neyman–Pearson lemma states that the likelihood ratio, $\Lambda$, given by:

\begin{equation}
    \Lambda (p_{\textrm{real}} \mid p_{\textrm{guess}})\equiv \frac {{\mathcal {L}}(z \mid p_{\textrm{real}}(z))}{{\mathcal {L}}(z \mid p_{\textrm{guess}}(z))} = \frac{p_{\textrm{real}}(z_1,z_2,...,z_n)}{p_{\textrm{guess}}(z_1,z_2,...,z_n)}
    \label{NP}
\end{equation}

is the most powerful test at the given significance level \cite{NeyPear}. Here, $p_{\textrm{real}}(z_i)$ and $p_{\textrm{guess}}(z_i)$ are the probabilities associated with the $i$-th decay in a jet having $n$ decays in total and following either the $p_{\textrm{real}}(z)$ or $p_{\textrm{guess}}(z)$ probability distributions.

This means that for a fixed $z$, if we find the most powerful test of distinguishing between jets created following the $p_{\textrm{real}}$ and $p_{\textrm{guess}}$ distributions, but we only know $p_{\textrm{guess}}$, we can recover $p_{\textrm{real}}(z)$. This can be done when several assumptions are satisfied. First of all, we consider that all the decays in a decay chain that produces a certain jet are independent. Hence, a jet can be described by a product of factors corresponding to the probability distribution as

\begin{equation}
    p(z_1,z_2,...,z_n) = p(z_1)p(z_2)...p(z_n) = p(z_1)p(z_2,...,z_n)\,, 
    \label{factorization}
\end{equation}

where $p(z_i)$ is the probability associated with a single decay in a jet having $n$ decays in total. Recall that in this notation $z_i$ is a set containing $z_{Ei}, z_{pi}, \phi_i$ and $\theta_i$ and the probability $p(z_i)$ can be written as $p(z_{E,i})p(z_{p,i}) p(\phi_i)p(\theta_i)$. Now let's select the same number of jets generated from $p_{\textrm{real}}(z)$ and $p_{\textrm{guess}}(z)$ and have a neural network distinguish between them. The neural network is set up as a classifier which gives the probability that the distribution generating an image is $p_{\textrm{real}}(z)$, i.e. it gives the value $C_{\textrm{nn}} \equiv p(p_{\textrm{real}} \mid z)$ \cite{NNProbability}. According to the Bayes' theorem, this value is equal to:

\begin{eqnarray}
p(p_{\textrm{real}} \mid z) &=& \frac{p(z \mid p_{\textrm{real}})p(p_{\textrm{real}})}{p(z \mid p_{\textrm{real}})p(p_{\textrm{real}})+p(z \mid p_{\textrm{guess}})p(p_{\textrm{guess}})} \nonumber \\
&=& \frac{p(z \mid p_{\textrm{real}})}{p(z \mid p_{\textrm{real}})+p(z \mid p_{\textrm{guess}})} = \frac{\Lambda(p_{\textrm{real}} \mid p_{\textrm{guess}})}{\Lambda(p_{\textrm{real}} \mid p_{\textrm{guess}})+1}\,,
\label{Bayes}
\end{eqnarray}

where we take into account the fact that $p(p_{\textrm{real}}) = p(p_{\textrm{guess}})$ since we take the same number of jet images from both distributions. By inverting (\ref{Bayes}) and using (\ref{NP}) and (\ref{factorization}), we obtain:

\begin{equation}
    \Lambda(p_{\textrm{real}} \mid p_{\textrm{guess}}) = \frac{C_{\textrm{nn}}}{1-C_{\textrm{nn}}} = \frac{p_{\textrm{real}}(z_1,z_2,...,z_n)}{p_{\textrm{guess}}(z_1,z_2,...,z_n)}=\frac{p_{\textrm{real}}(z_1)p_{\textrm{real}}(z_2,...,z_n)}{p_{\textrm{guess}}(z_1)p_{\textrm{guess}}(z_2,...,z_n)}\,.
    \label{likelihood}
\end{equation}

Now let's look at only $p_{\textrm{real}}(z_1)$, i.e. the real probability distribution, but for a fixed $z_1$. An inversion of (\ref{likelihood}) gives:

\begin{equation}
    p_{\textrm{real}}(z_1) = \frac{C_{\textrm{nn}}}{1-C_{\textrm{nn}}}\cdot  p_{\textrm{guess}}(z_1)\cdot \frac{p_{\textrm{guess}}(z_2,...,z_n)}{p_{\textrm{real}}(z_2,...,z_n)}\,.
    \label{p_real_final-1}
\end{equation}

In our case, this can be applied to $p_E(z_{E,1})$ and $p_p(z_{p,1})$ distributions:

\begin{eqnarray}
    p_{\textrm{real},E}(z_{E,1}) &=& \frac{C_{\textrm{nn}}}{1-C_{\textrm{nn}}}\cdot p_{\textrm{guess},E}(z_{E,1})\cdot \frac{p_{\textrm{guess},p}(z_{p,1})p_{\textrm{guess}}(z_2,...,z_n)}{p_{\textrm{real},p}(z_{p,1})p_{\textrm{real}}(z_2,...,z_n)}\hskip 3mm \textrm{and} \nonumber \\
    p_{\textrm{real},p}(z_{p,1}) &=& \frac{C_{\textrm{nn}}}{1-C_{\textrm{nn}}}\cdot p_{\textrm{guess},p}(z_{p,1})\cdot \frac{p_{\textrm{guess},E}(z_{E,1})p_{\textrm{guess}}(z_2,...,z_n)}{p_{\textrm{real},p}(z_{E,1})p_{\textrm{real}}(z_2,...,z_n)}\,.
    \label{p_real_final}
\end{eqnarray}

This final expression offers a possibility of recovering $p_{\textrm{real, E}}$ and $p_{\textrm{real, p}}$ by only knowing $p_{\textrm{guess,E}}$ and $p_{\textrm{guess,p}}$ in the case the neural network acts as an ideal classifier. It is assumed that all of the angles occur with equal probabilities so they are omitted from the equation. A detailed description of how expression is implemented is given in the Methods and Materials section.

\subsection{Recovering the original probability distribution}

In what follows discussion, the indices $E$ and $p$ are omitted to improve clarity, but the general conclusions work for either the energy distribution $p_{\textrm{guess,E}}$ or the momentum distribution  $p_{\textrm{guess,p}}$. To provide a reasonable $p_{\textrm{guess}}$ distribution, we have to know some of the background of the physical process that governs $p_{\textrm{real}}$. For example, from our physics background we know that this distribution should fall with increasing $z$. An example of such a distribution is 

\begin{equation}
    p_{\textrm{guess}}(z) = \mathcal{N}\textrm{e}^{-Cz} \,,
    \label{eq_probna}
\end{equation}

whose integral is normalized to 1. This distribution is allowed to be only ``good enough'' when using the outlined procedure, since we can iteratively repeat it and set

\begin{equation}
    p^{i+1}_{\textrm{guess}}(z) = p^{i}_{\textrm{real, calculated}}(z)\,,
    \label{eq_iterate}
\end{equation}

with $i$ being the iteration index and $p^{i}_{\textrm{real, calculated}}(z)$ being the approximation of the ``real'' distribution as determined in the current step. The reason why the guess distribution converges to the real distribution when applying this procedure iteratively can be seen if one looks at the cross entropy loss of the neural network. This quantity, also known as the log loss, measures the performance of a classification model where the prediction input is a probability value between 0 and 1 \cite{CrossEntropy}. In the case of binary classification, which we perform here, and using the notation already given in the text, it is given by:

\begin{equation}
L = -\frac{1}{2}\sum_{i=1}^{n}\left[y(z_i)\log C_{\textrm{nn}} + (1-y(z_i))\log (1-C_{\textrm{nn}}) \right]\,.
\end{equation}

Here, $y(z_i)$ is the set of true data labels, being either 1 or 0, depending on which distribution was used to create a particular jet. In general, the performance of any model is always worse compared to the ideal model, so that the cross entropy loss of our classifier $L$ has to be larger than the loss of an ideal classifier $L_{\textrm{ideal}}$. Using (\ref{Bayes}), this can be written as:

\begin{eqnarray}
L &>& -\frac{1}{2}\sum_{i=1}^{n}\left[y(z_i)\log C_{\textrm{nn}}^{\textrm{ideal}} + (1-y(z_i))\log (1-C_{\textrm{nn}}^{\textrm{ideal}}) \right] \nonumber \\
&>& -\frac{1}{2}\sum_{i=1}^{n}\left[y(z_i)\log\left( \frac{\Lambda(p_{\textrm{real}} \mid p_{\textrm{guess}})}{1+\Lambda(p_{\textrm{real}} \mid p_{\textrm{guess}})}\right) + (1-y(z_i))\log \left( \frac{1}{1+\Lambda(p_{\textrm{real}} \mid p_{\textrm{guess}})}\right) \right]
\label{LogLoss}
\end{eqnarray}

Now let's assume that the classifiers have been fed only the data from the real distribution, i.e. that we set $p^{i}_{\textrm{guess}} = p_{\textrm{real}}$ on purpose. Then the data labels $y(z_i)$ are all equal to 1, so that

\begin{equation}
-\frac{1}{2}\sum_{i=1}^{n}\log C_{\textrm{nn}} > -\frac{1}{2}\sum_{i=1}^{n}\left[\log\left( \frac{\Lambda(p_{\textrm{real}} \mid p_{\textrm{guess}})}{1+\Lambda(p_{\textrm{real}} \mid p_{\textrm{guess}})}\right) \right]\,.
\end{equation}

Although the index $i$ has been left out to improve readability, both expressions under the sum still depend on the selected $z$-bin. A short rearrangement of this condition gives:

\begin{equation}
\prod_{i=1}^{n} C_{\textrm{nn}} < \prod_{i=1}^{n} \frac{\Lambda(p_{\textrm{real}} \mid p_{\textrm{guess}})}{1+\Lambda(p_{\textrm{real}} \mid p_{\textrm{guess}})}\,.
\end{equation}

Now we use the fact that $C_{\textrm{nn}} > 0.5$, which we know to be true averaged over $z$, if the network has any discriminating power. Using (\ref{Bayes}) again, the last inequality can be rearranged into:

\begin{equation}
\prod_{i=1}^{n} \frac{C_{\textrm{nn}}}{1-C_{\textrm{nn}}} < \prod_{i=1}^{n} \Lambda(p_{\textrm{real}} \mid p_{\textrm{guess}})\,,
\end{equation}

Note that the left side of this inequality is a product larger than one, even though some of the factors after the product sign can be smaller than one. Recalling the definition of $\Lambda(p_{\textrm{real}} \mid p_{\textrm{guess}})$, after multiplying with $p_{\textrm{guess}}$ we can write:

\begin{equation}
\prod_{i=1}^{n} p_{\textrm{guess}}(z_i) <
\prod_{i=1}^{n} \frac{C_{\textrm{nn}}}{1-C_{\textrm{nn}}} p_{\textrm{guess}}(z_i) < \prod_{i=1}^{n} p_{\textrm{real}}(z_i)\,.
\end{equation}

The first term on the left is the guess distribution in one of the iterations, the second term is the next iteration of the guess distribution since we are using (\ref{p_real_final}) and (\ref{eq_iterate}) and the last term is the real distribution. Thus we can conclude that in this case, the iterations successively converge to the real distribution. The same argument can be used when the network is fed only the data from the guess distribution. Since the real data is a mix of the two we conclude that in general, the successive iterations of the guess distribution converge on average to the real distribution. If we could perform an infinite number of iterations, we would reach the real distribution from the guess distribution, but since we have limited time and resources, the two will always be at least slightly different.

\subsection{Calculation results and errors}

Our calculation was performed with the initial guess probability distributions given by $p_{\textrm{guess,E}} = p_{E}^0(z_E) =  \mathcal{N}_E\textrm{e}^{-Cz_E}$, with $z_E$ in the interval $[0.01, 0.5]$ and $p_{\textrm{guess,p}}=p_{p}^0(z_p) =  \mathcal{N}_p\textrm{e}^{-Cz_p}$ with $z_p$ in the interval $[0.01,1]$. Three different values of the constant $C$ were used, 0.1, 10 and 100,  thus creating a nearly flat distribution, a distribution slowly decreasing with increasing $z$ and a much more rapidly decreasing distribution, respectively.   

Once the iterative procedure starts, we need to decide at which point to stop further iterations. We define the error margin of the $i$-th iteration for a single variable (either energy or momentum) of a guess distribution as the root mean square relative error (RMSRE), which is a typical cross-validation tool \cite{RMSRE}:

\begin{equation}
    \textrm{RMSRE} = \sqrt{\frac{1}{10}\sum_{j=1}^{10} \left(1-\frac{p^{i}_{\textrm{guess}}(z_j)}{p_{\textrm{real}}(z_j)}\right)^2}\,.
    \label{eq_error}
\end{equation}

The index $j$ comes from the fact that we had to choose a number of $z$ bins, which we set to 10, in order to perform the calculations. The total error margin for an iteration of a guess distribution is defined as the arithmetic mean of the margins for energy and momentum. We stop the iterative procedure once the error margin remains below 10\% during 20 successive iterations \cite{ErrorMargin}.

\begin{figure}[h!t!]
\centering
\includegraphics[width=15cm]{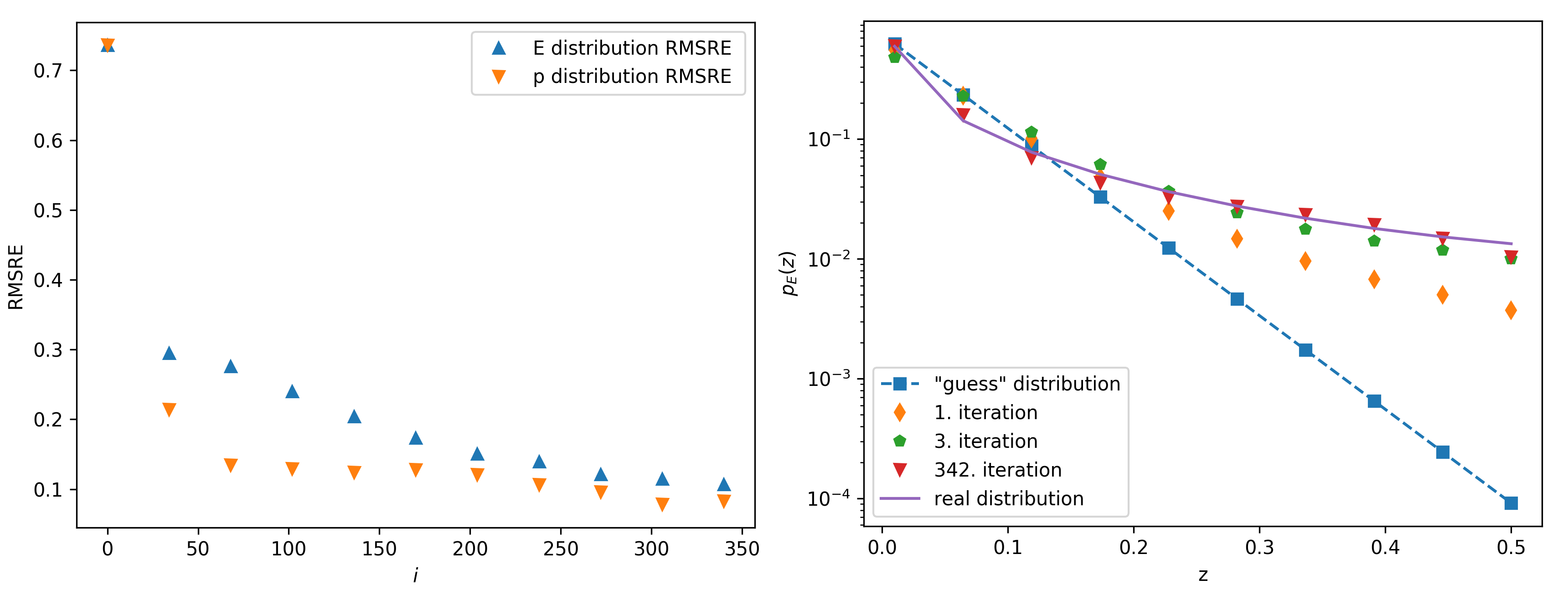}
\caption{(\textbf{a}) The calculated error margin vs. the iteration number in the case of the guess distribution given by (\ref{eq_probna}) with $C$ set to 10. The error calculation is described in the text. The error margins are shown separately for the case when the classifier is trained with jet images populated either with jet energies or jet momenta. (\textbf{b}) Several iterations of the calculated probability distributions $p_E^i(z)$ (symbols) compared to $p_{\textrm{real,E}}(z)$ (full line). The 342$^{\textrm{nd}}$ iteration is the final iteration of this procedure, since the stopping condition has been satisfied.}
\label{image_error_10}
\end{figure}

A graph showing the dependence of the error margin on the iteration index for the case of distribution (\ref{eq_probna}) with $C$ set to 10 is given in Fig.(\ref{image_error_10}). The graph shows the margins for the calculated $p^{i}_{E}(z)$ and $p^{i}_{p}(z)$ distributions. In this case the average error margin is lower than 10\% for 20 successive iterations after the 342$^{\textrm{nd}}$ iteration. On the same figure we also show the calculated probability distributions $p^{i}_{E}(z)$ and compare them to $p_{\textrm{real},E}(z)$. The comparison of the probability distributions $p^{i}_{p}(z)$ to $p_{\textrm{real},p}(z)$ for different parameters $C$ is given in Appendix B.

\begin{figure}[h!t!]
\centering
\includegraphics[width=14cm]{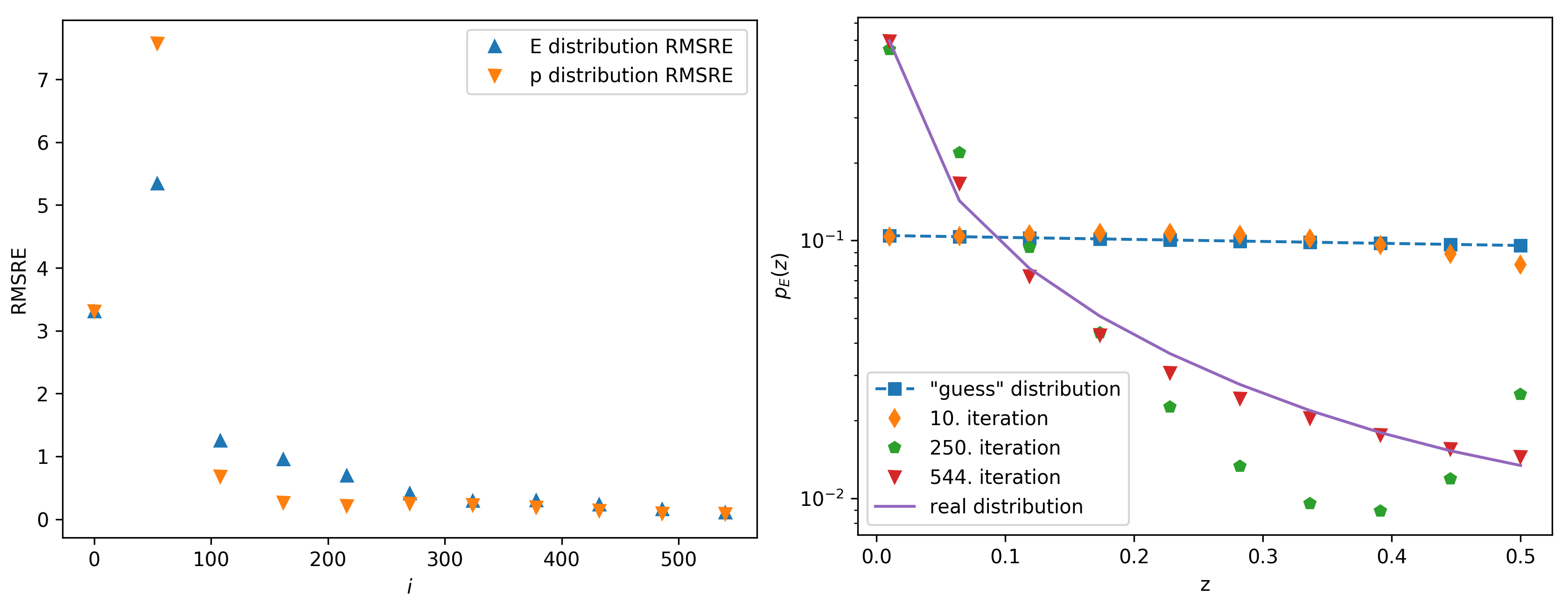}
\caption{(\textbf{a}) The calculated error margin vs. the iteration number in the case of the guess distribution given by (\ref{eq_probna}) with $C$ set to 0.1. (\textbf{b}) Several iterations of the calculated probability distributions $p_E^i(z)$ (symbols) compared to $p_{\textrm{real, E}}(z)$ (full line). The 544$^{\textrm{th}}$ iteration is the final iteration of this procedure, since the stopping condition has been satisfied.}
\label{image_error_01}
\end{figure}

\begin{figure}[h!t!]
\centering
\includegraphics[width=15cm]{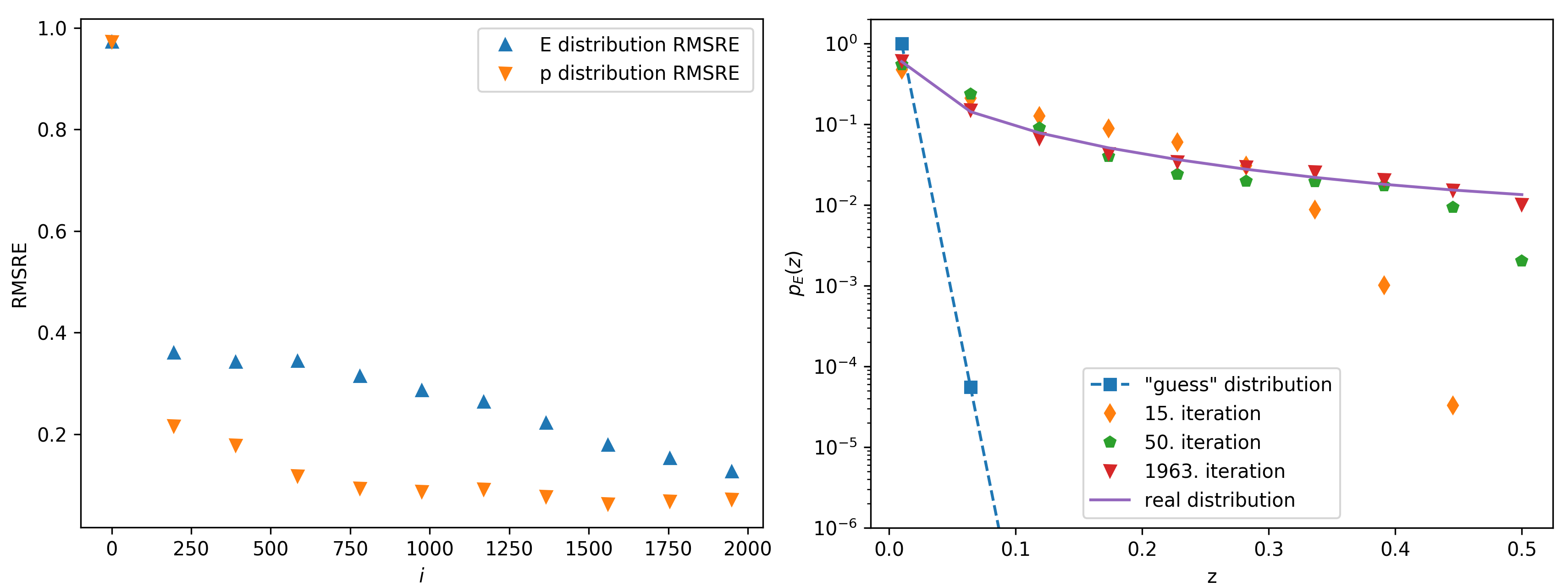}
\caption{(\textbf{a}) The calculated error margin vs. the iteration number in the case of the guess distribution given by (\ref{eq_probna}) with $C$ set to 100. (\textbf{b}) Several iterations of the calculated probability distributions $p_E^i(z)$ (symbols) compared to $p_{\textrm{real,E}}(z)$ (full line). The 1963$^{\textrm{rd}}$ iteration is the final iteration of this procedure, since the stopping condition has been satisfied.}
\label{image_error_100}
\end{figure}

One can see the decrease of the error margin with growing iteration index and the convergence of the guess distribution to the real distribution. The graphs showing the dependence of the error margins on the iteration index and the calculated probability distributions  $p^{i}_{E}(z)$ compared to $p_{\textrm{real,E}}(z)$ when $C$ equals 0.1 and 100, respectively, are given in Figs. (\ref{image_error_01}) and (\ref{image_error_100}). In these cases, the stopping condition has been reached after 544 and 1963 iterations, respectively. When comparing the results for different initial guess distributions, we note that only the total number of the iterations needed to achieve sufficient convergence is affected by the initial conditions.

\section{Discussion}

In this paper we present a study performed on a toy model representing a simplified version of a QCD fragmentation process. It is possible to retrieve some of the unknown properties of this process by using a correct interpretation of a neural network model combined with incomplete knowledge of the system. We presented an iterative method which recovers unknown probability distributions that govern the presented physical system. We have mathematically shown that one can expect the convergence from our incomplete knowledge to the real underlying distributions by using the developed method. This claim was confirmed by our results. 

The method we chose requires an initial guess of the probability distributions from which the original distributions are to be recovered. The choice of the guessed probability distributions affects only the number of iterations needed to achieve the convergence to the real distributions. The final error margin between the obtained distributions and the real distributions should depend only on the discriminating power of the used classifier, i.e. the convolutional neural network. In our study, we used a stopping condition which relies on the RMSRE between the real and the calculated distribution. However, this relies on the fact that we constructed and knew the real distribution, which is not true in a realistic setting. In that case, the stopping condition could be based solely on the the loss function of the classifier, evaluated on some test dataset. For example, one could impose the condition that the values of the loss function are in some small interval around the minimal possible loss value $L_{\textrm{min}}$. In that case the expected values of the classifier output will lie in some small interval around $C_{\textrm{nn}} = 0.5$ and any further calculation will not significantly improve the probability distributions obtained in the previous iteration. 

Since this method doesn't imply what kind of classifier should be used, any machine learning technique used for binary classification can be employed. In this research we developed a classifier based on a  convolutional neural networks, which have proven to be very successful in the image classification tasks. We believe that the presented method can be generalized for use in more realistic physical systems which include multiple decay mechanisms. This would improve the similarity to the real QCD fragmentation, where a quark can radiate a gluon or vice versa and quark-antiquark pairs can be formed from gluons. In such a scenario, this method could be applied to real data collected by some high energy experiment.

\section{Materials and Methods}

In this section we present in detail the methodology used to obtain the presented results. First, we describe the jet generator used to crate the jet images. Next, we present the detailed architecture of the neural network used as the classifier and finally, we detail the algorithm used to recover of the underlying probability distributions.

The computational code used to develop the particle generator, the neural network model and the calculation of the probability distributions is written in the the Python programming language using the Keras module with the TensorFlow backend \cite{keras}. Both the classifier training and jet generating were performed using a standardized PC setup equiped with an NVIDIA Quadro p6000 graphics processing unit.

\subsection{The jet generator}

\begin{enumerate}[leftmargin=*,labelsep=4.9mm]

\item We start with a particle at rest with a given rest mass, here taken to be $m_0=\,$100 (the units are inconsequential in the calculation).
\item The particle decays into two new particles. The energies and the momenta of these particles are determined by a probability distribution. To generate the ``real'' data we use a distribution already known in particle physics, given by:
    
\begin{equation}
p(z) = \mathcal{N}\frac{1+(1-z)^2}{z}\,.
\label{eq_stvarna}
\end{equation}

The energy of the decay particle $E$ equals $zE_0$, with $E_0=m_0$ being the energy of the decaying particle. Note that the probability diverges as $z$ approaches zero, so the distribution is limited by a lower boundary on $z$ both due to physical and computational reasons. $\mathcal{N}$ is a constant that ensures that the integral of the probability distribution equals 1 and depends on the lower boundary set on $z$. In our simulation, we set the minimum $z$ to 10$^{-2}$, making $\mathcal{N}$ equal to $\approx 0.13$. 

The momentum of the decay particle is limited with the total energy of the particle. We determine the momentum by sampling the same probability distribution as for the energy, but now we set the momentum $p$ equal to $zE$, with $E$ being the energy of the decay particle. To differentiate between these $z$ distributions, we write $z_E$ and $z_p$ when deemed necessary.

The spatial distribution of the decay products is uniform in space. This means that, observed from the rest frame of the decaying particle, the probability that either one of the decay products flies off in a certain infinitesimal solid angle is uniform. Physically speaking, the angles $\theta$ and $\phi$ are sampled from uniform distributions on intervals $[0, \pi]$ and $[0, 2\pi]$ respectively.

The energy, the momentum and the direction of the second particle are determined by the laws of conservation of energy and momentum. In other words, $z_1 + z_2 = 1$ when looking at energy, and $p_1 + p_2 = 0$, since the original momentum in the center of mass system is zero. These facts also save computational time due to symmetry, since we can sample for the energy of the first particle in the interval $\left[0.01,0.5\right]$, instead of placing the upper limit for $z$ to 1.

\item After the first decay, the procedure repeats iteratively, i.e. we repeat step 2 for both decay products from the previous step. The only difference compared to the previous step is that we now perform the calculations for each particle in its center of mass frame and then transform the obtained quantities back to the laboratory frame, which coincides with the center of mass frame of the original particle.

Once the total number of particles exceeds a pre-determined threshold (in our case set to 32), we disregard the lowest energy particles. We do this both to reduce the computational time and because we determined that these particles don't influence our end result in a significant manner.

The decay procedure stops when either of two conditions is met; if the decay particle mass falls below 0.1, or a certain number of decays has been reached. In the simulations, we limited the number of decays in a single branch to 50. For simplicity, all the decays are considered to happen in the same point in space.

\item The list of final decay particles now forms a list that contains the energies, the momenta and the directions of the $n$ particles. We call this entity a jet. The jet has a maximum of 32 particles in its final state stemming from a maximum of 1 + 2 + 4 + 8 + 16 + 45$\cdot$32 = 1471 decays. Hence, the full description of a jet is given by a maximum of 1471 $z_E$ parameters, 1471 $z_p$ parameters and 1471 pairs of angles ($\theta$, $\phi$). 

To create the final representation of the jet which will be fed to a classifier we create a histogram whose axes represent the direction of a particle in space. The histogram has 32$\times$32 pixels with axes representing the polar angle $\theta$ and the azimuthal angle $\phi$ of a particle. The color of a pixel in the histogram corresponds to either the energy or the momentum of the particle traveling in that direction in space. We distribute the deposited energy and momentum as Gaussian distributions in the histograms, with the Gaussian of $\sigma$ equal to 1 pixel centralized at the pixel corresponding to a direction of a certain particle. This mimics the physics situation in real life, where the readout from a detector always consists of a signal and a background noise. In fact, even when simulating data in a deterministic way, this effect is taken into account \cite{geant}. Lastly, the energy and momentum histograms are stacked to create an image with dimensions 32$\times$32$\times$2. An example of the jet generator tree with modified parameters is given in the appendix. Two examples of jet images are given on figure (\ref{JetImage}) in the main body of the text.
\end{enumerate}

\subsection{The classifier}

The classifier used to recover the real probability distribution is a feed forward convolutional neural network (CNN) \cite{cnn}. The architecture of the used CNN is schematically shown on Fig. \ref{Algorithms}. It consists of a block of layers, repeated four times, followed by 3 dense layers consisting of 20, 10 and 1 unit respectively. A ReLu activation function is used in each layer, except for the last one, where a sigmoid function is used. The layer block consists of a 2-dimensional convolutional layer (with 32 filters and a (3,3) kernel), a MaxPooling layer, a batch normalization layer and a dropout layer. The training of the classifier is performed by minimizing the binary cross entropy loss \cite{CrossEntropy}. The AdaM optimizer is used to optimize the weights of the CNN \cite{adam}. When training through the iterations, in each iteration we use the same number of jets obtained with $p_{real}(z)$ and jets obtained by using the distribution calculated from the previous iteration. To train the CNN we use 75\% of data, while the remaining 25\% are used to validate the trained model.

\subsection{The algorithm used to recover the underlying probability distributions}

\begin{figure}[h!t!]
\centering
\resizebox{0.65\textwidth}{!}{
\begin{tikzpicture}[scale=1.0,transform shape]
  \draw (-3,1.2) -- ++(12,0);
  \draw (2.8,2) -- ++(0,-17);
  \node[] at (0,1.5) {\large Layers};  
  \node[] at (5.8,1.5) {\large Output dimensions}; 

  \path \inputblock {1}{};
  \path (inp1.south)+(0.0,-1.5) \block{1}{Conv 2D};
  \path (inp1.east)+(4.0,0.0) \bldimension{1}{32x32x2};
  
  \path (p1.south)+(0.0,-1.0) \block{2}{MaxPooling};
  \path (p2.east)+(4.0,-0.7) \bldimension{2}{1. pass\\ 16x16x32 \vskip 1mm 2. pass\\ 8x8x32 \vskip 1mm 3. pass\\4x4x32 \vskip 1mm 4. pass\\ 2x2x32};
  
  \path (p2.south)+(0.0,-1.0) \block{3}{BatchNorm};
  \path (p3.south)+(0.0,-1.0) \block{4}{Dropout};
  \path (p4.south)+(0.0,-1.5) \flattenblock{1}{};
  \path (flb1.south)+(0.0,-1.5) \denseblock{1}{};
  \path (flb1.east)+(4.0,0.0) \bldimension{6}{128};
  \path (dnb1.south)+(0.0,-1) \denseblock{2}{};
  \path (dnb1.east)+(4.0,0.0) \bldimension{7}{20};
  \path (dnb2.south)+(0.0,-1) \denseblock{3}{};
  \path (dnb2.east)+(4.0,0.0) \bldimension{8}{10};
  \path (dnb3.east)+(4.0,0.0) \bldimension{9}{1};
  
  \path [line] (inp1.south) -- node [above] {} (p1);        
  \path [line] (p1.south) -- node [above] {} (p2);
  \path [line] (p2.south) -- node [above] {} (p3);
  \path [line] (p3.south) -- node [above] {} (p4);
  \path [line] (p4.south) -- node [right] {\textbf{5$^{\textrm{th}}$ pass}} (flb1);
  
  \path [line] (p4.west) -- ++(-2,0) -- ++(0,4.6)-- node [above] {\textbf{4 passes}} (p1.west);
  
  \path [line] (flb1.south) -- node [above] {} (dnb1);
  \path [line] (dnb1.south) -- node [above] {} (dnb2);
  \path [line] (dnb2.south) -- node [above] {} (dnb3);
   
  \background{p2}{p1}{p3}{p4}{Blocks}
  \background{bld2}{bld2}{bld2}{bld2}{Blocks}
  \background{dnb2}{dnb1}{dnb2}{dnb3}{Dense}
\end{tikzpicture}%
}
\hskip -1mm
\resizebox{0.3\textwidth}{!}{
\raisebox{10em}{\begin{tikzpicture}[scale=1.0,transform shape]
  \path \block {1}{Generate data from $p_{i-1}$};
  \path (p1.south)+(0.0,-1.8) \block{2}{Train the classifier\\ 7500 real\\ 7500 $p_{i-1}$\\ 5000 data};
  \path (p2.south)+(0.0,-1.8) \block{3}{Calculate the distribution:\\
      1. Generate\\
      2. Calculate\\
      3. Fit};
  \path (p3.south)+(0.0,-1.8) \block{4}{Check condition};
  \path [line] (p1.south) -- node [above] {} (p2);        
  \path [line] (p2.south) -- node [above] {} (p3);
  \path [line] (p3.south) -- node [above] {} (p4);
  \path [line] (p4.east) node [right, yshift=0.25cm] {\textbf{If AUC $<$ 0.5}} -- ++(2.5,0) -- ++(0,8.5)--(p1.east);
\end{tikzpicture}%
}}
\caption{(\textbf{a}) The left panel shows the architecture of the convolutional neural network as described in the text. The output dimensions of each layer are given on the right side of the panel. The Blocks layer goes through 4 passes. (\textbf{b}) The right panel shows the algorithm used to recover the underlying probability distributions. AUC stands for Area Under the Curve and provides an aggregate measure of the network performance.}
\label{Algorithms}
\end{figure}
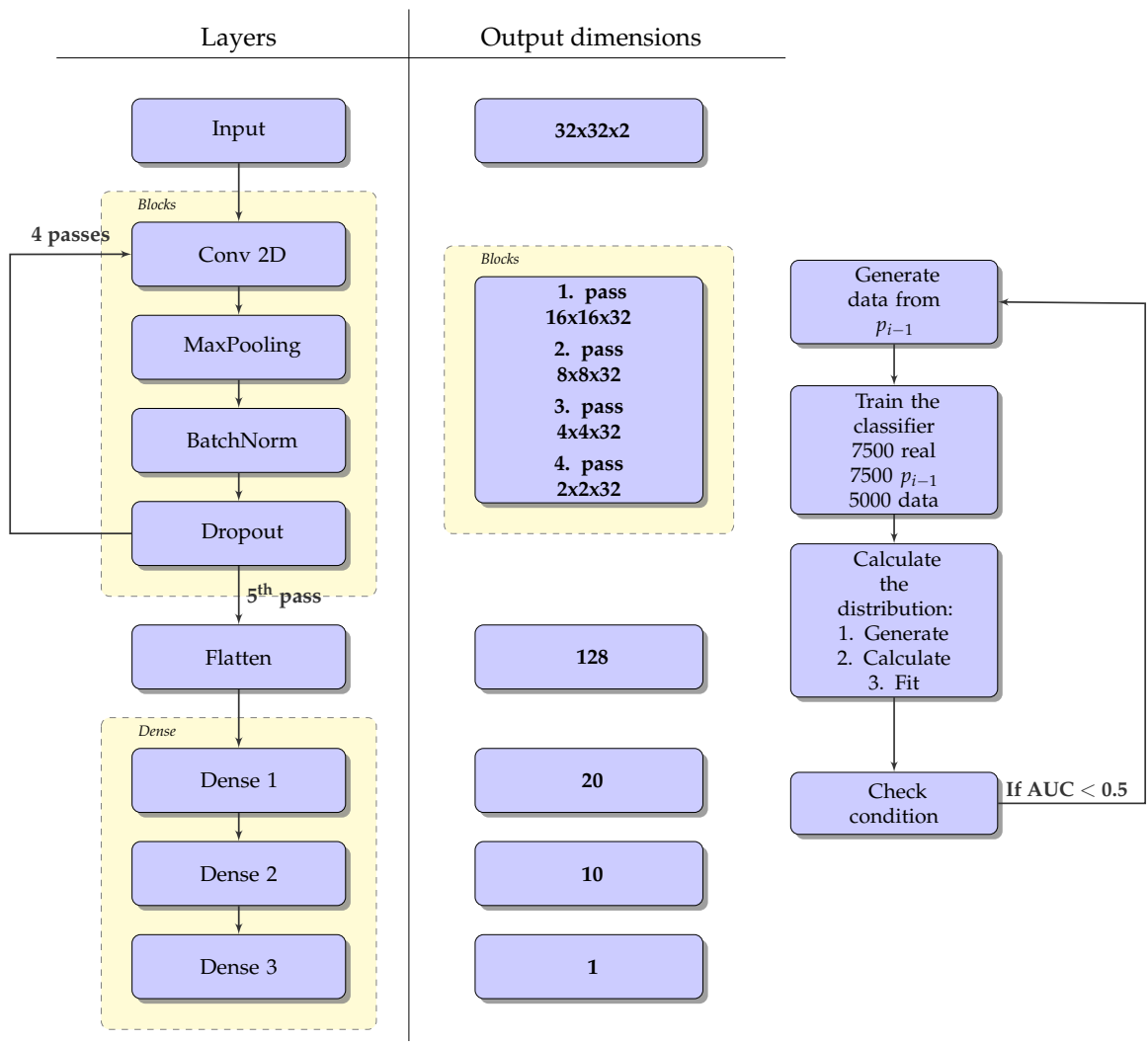

After the ``real'' jets dataset is generated, we want to recover its underlying probability distributions, which we treat as unknown. The schematic view of the algorithm we use for this is shown on Fig. \ref{Algorithms}. The algorithm is repeated iteratively. To begin with, we set some initial guesses of the underlying distributions, denoted by $p_E^0(z_E)$ and $p_p^0(z_p)$. Each iteration, indexed by $i$, consists of 3 steps: first, the data is generated using the probability distributions $p_E^i(z_E)$ and $p_p^i(z_p)$. Next, the classifier is trained on the generated data and then the probability distributions $p_E^{i+1}(z_E)$ and $p_p^{i+1}(z_p)$ are calculated using the trained classifier. After each iteration the weights of the classifier are saved and used as the initial weights for the training procedure in the next iteration. The iterative procedure is stopped once the error margin (\ref{eq_error}) remains below 10\% during 20 successive iterations. Further subsections present the details of the outlined algorithm.

\subsubsection{Generating the data from the obtained distributions}

To generate the data used for the next iteration we sample 10 000 vectors $z \equiv (z_1,z_2,...,z_N)$, where $N_{\textrm{max}} = 1471$ and $z_n \equiv (z_{E}^n, z_{p}^n, \theta^n, \phi^n)$. The parameters $z_E$ and $z_p$ are sampled from the $p_E^i(z_E)$ and $p_p^{i}(z_p)$ probability distributions obtained from the previous iteration. These vectors are fed into the jet generator to obtain 10 000 jet images.  

\subsubsection{Training the CNN classifier}

The generated data is next used to train the classifier. The 10 000 samples of jets generated by the distributions $p_E^i(z_E)$ and $p_p^i(z_p)$ are paired with 10 000 randomly chosen samples from the dataset containing jets generated using the ``real'' distributions. If the $p_E^i(z_E)$ and $p_p^i(z_p)$ distributions and the ``real'' distributions are very different, the output of the classifier $C_{nn}$ can be expected to be very close to 0 or 1. This can occur during the early iterations of the algorithm and can cause computational difficulties due to nature of the denominator in (\ref{p_real_final}). To avoid these difficulties, the classifier is trained on a smaller dataset during the early iterations, typically containing $\approx$ 200-2000 jets.  

\subsubsection{Calculation of the probability distributions}

In order to calculate $p_E^{i+1}(z_E)$ and $p_p^{i+1}(z_p)$, we use (\ref{p_real_final}). First, we generate a vector $z \equiv (z_1,z_2,...,z_N)$, where $z_n = (z_{E}^n, z_{p}^n, \theta^n, \phi^n)$, by sampling the $p_E^i(Z_E)$ and $p_p^{i}(z_p)$ probability distributions. From each of these vectors we remove $z_{E}^1$ and fix it by hand to a value between 0.01 and 0.5 in 1000 equidistant bins. This way, we create 1000 vectors $z$ which differ only in the $z_E$ parameter of the first decay. Our jet generator is then used to create the jet images. Each of the images is fed to the classifier, which gives us $C^{j}_{\textrm{nn}}(z_j)$, with $j$ being the index of the image. The second term in (\ref{p_real_final}) can be directly calculated using the $p_{E}^i(z_{E}^i)$ distribution. The last two terms form a constant which is equal for all of the used jet images. The calculation of the constant is simple: since we are dealing with probability distributions, we impose the condition that the integral $p_E(z_E)$ over $z_E$ equals 1, which directly determines the value of the constant. This way, we obtain the value of the probability distribution $p_E(z_{E})$ for each $z_E$ bin. This procedure is repeated 200 times with jet images generated with different decay conditions. The arithmetic average of the calculated distributions is used to finally determine the distribution $p_{E}^{i+1}$. Due to the nature of the algorithm, $p^{i}_E(z_E)$ inevitably won't be a smooth function since it is calculated on a point to point basis. Before feeding this distribution to the next iteration, we perform a smoothing by fitting a fourth degree polynomial to the calculated values on a log scale. An analogous procedure is used to determine $p_p^i(z_p)$. The only difference is that instead of $z_E$ in this case we fix the $z_p$ parameter of the first decay.  

\vspace{6pt} 

\authorcontributions{Both authors contributed equally to all segments of this paper.}

\funding{This research was funded by the Croatian science foundation grant IP-2018-01-4108 ``Demystifying Two Particle Correlations in pp collisions with the upgraded Time Projection Chamber''.}

\acknowledgments{We gratefully acknowledge the support of NVIDIA Corporation with the donation of the Quadro P6000 graphics processing unit used for this research.}

\conflictsofinterest{The authors declare no conflict of interest.} 

\abbreviations{The following abbreviations are used in this manuscript:\\
\noindent 
\begin{tabular}{@{}ll}
QCD & Quantum Chromodynamics\\
LHC & Large Hadron Collider\\
RMSRE & Root mean square relative error\\
CNN & Convolutional Neural Network\\
AUC & Area Under the Curve
\end{tabular}}

\newpage
\appendixtitles{yes} 

\appendix
\section{An example of a generated jet}
Here we give a pictorial example of a jet generated as outlined in section 2.

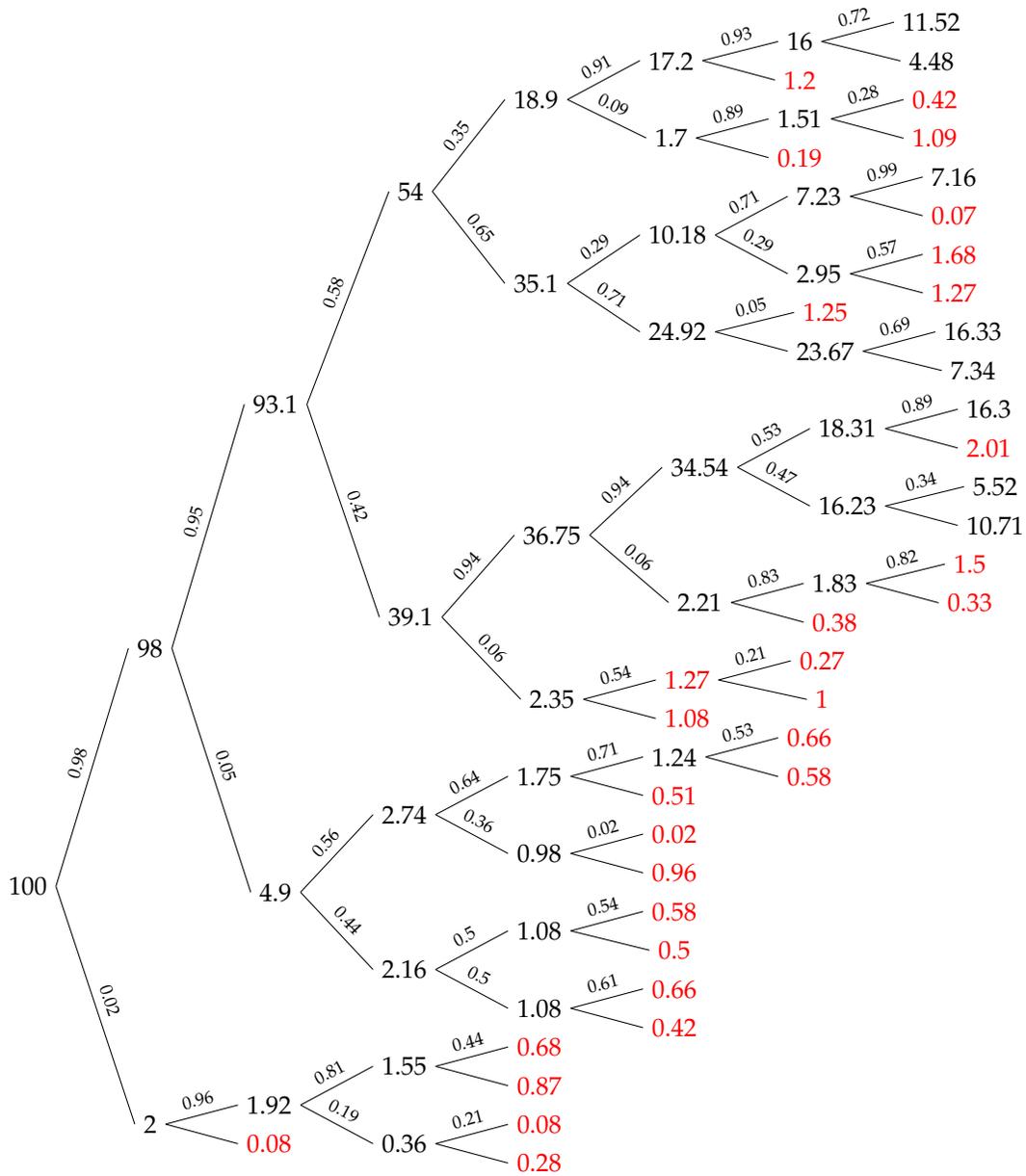
\begin{figure}[h!t!]
  \centering   
\begin{forest}
for tree={
      grow=east,
      parent anchor=east,
      child anchor=west,
      s sep=1pt,
      l sep=1cm
     },
[100
 [2,edge label={node[midway,above, sloped,font=\scriptsize]{0.02}}
  [0.08,text=red,edge label={node[midway,above, sloped,font=\scriptsize]{}}]
   [1.92,edge label={node[midway,above, sloped,font=\scriptsize]{0.96}}
    [0.36,edge label={node[midway,above, sloped,font=\scriptsize]{0.19}}
     [0.28,text=red, edge label={node[midway,above, sloped,font=\scriptsize]{}}]
     [0.08,text=red, edge label={node[midway,above, sloped,font=\scriptsize]{0.21}}]
    ]
    [1.55,edge label={node[midway,above, sloped,font=\scriptsize]{0.81}}
     [0.87,text=red,edge label={node[midway,above, sloped,font=\scriptsize]{}}]
     [0.68,text=red, edge label={node[midway,above, sloped,font=\scriptsize]{0.44}}]
    ]
   ]
  ]
  [98,edge label={node[midway,above, sloped,font=\scriptsize]{0.98}}
   [4.9,edge label={node[midway,above, sloped,font=\scriptsize]{0.05}}
    [2.16,edge label={node[midway,above, sloped,font=\scriptsize]{0.44}}
     [1.08,edge label={node[midway,above, sloped,font=\scriptsize]{0.5}}
      [0.42,text=red, edge label={node[midway,above, sloped,font=\scriptsize]{}}]
      [0.66,text=red, edge label={node[midway,above, sloped,font=\scriptsize]{0.61}}]
     ]
     [1.08,edge label={node[midway,above, sloped,font=\scriptsize]{0.5}}
      [0.5,text=red, edge label={node[midway,above, sloped,font=\scriptsize]{}}]
      [0.58,text=red, edge label={node[midway,above, sloped,font=\scriptsize]{0.54}}]
     ]
    ]
    [2.74,edge label={node[midway,above, sloped,font=\scriptsize]{0.56}}
     [0.98,edge label={node[midway,above, sloped,font=\scriptsize]{0.36}}
      [0.96,text=red, edge label={node[midway,above, sloped,font=\scriptsize]{}}]
      [0.02,text=red,edge label={node[midway,above, sloped,font=\scriptsize]{0.02}}]
     ]
     [1.75,edge label={node[midway,above, sloped,font=\scriptsize]{0.64}}
      [0.51,text=red, edge label={node[midway,above, sloped,font=\scriptsize]{}}]
      [1.24,edge label={node[midway,above, sloped,font=\scriptsize]{0.71}}
       [0.58,text=red, edge label={node[midway,above, sloped,font=\scriptsize]{}}]
       [0.66,text=red, edge label={node[midway,above, sloped,font=\scriptsize]{0.53}}]
      ]
     ]
    ]
   ]
   [93.1,edge label={node[midway,above, sloped,font=\scriptsize]{0.95}}
    [39.1,edge label={node[midway,above, sloped,font=\scriptsize]{0.42}}
     [2.35,edge label={node[midway,above, sloped,font=\scriptsize]{0.06}}
      [1.08,text=red, edge label={node[midway,above, sloped,font=\scriptsize]{}}]
      [1.27,text=red, edge label={node[midway,above, sloped,font=\scriptsize]{0.54}}
       [1,text=red, edge label={node[midway,above, sloped,font=\scriptsize]{}}]
       [0.27,text=red, edge label={node[midway,above, sloped,font=\scriptsize]{0.21}}]
      ]
     ]
     [36.75,edge label={node[midway,above, sloped,font=\scriptsize]{0.94}}
      [2.21,edge label={node[midway,above, sloped,font=\scriptsize]{0.06}}
       [0.38,text=red, edge label={node[midway,above, sloped,font=\scriptsize]{}}]
       [1.83,edge label={node[midway,above, sloped,font=\scriptsize]{0.83}}
        [0.33,text=red, edge label={node[midway,above, sloped,font=\scriptsize]{}}]
        [1.5,text=red,edge label={node[midway,above, sloped,font=\scriptsize]{0.82}}]
       ]
      ]
      [34.54,edge label={node[midway,above, sloped,font=\scriptsize]{0.94}}
       [16.23,edge label={node[midway,above, sloped,font=\scriptsize]{0.47}}
        [10.71,edge label={node[midway,above, sloped,font=\scriptsize]{}}]
        [5.52,edge label={node[midway,above, sloped,font=\scriptsize]{0.34}}]
       ]
       [18.31,edge label={node[midway,above, sloped,font=\scriptsize]{0.53}}
        [2.01,text=red, edge label={node[midway,above, sloped,font=\scriptsize]{}}]
        [16.3,edge label={node[midway,above, sloped,font=\scriptsize]{0.89}}]
       ]
      ]
     ]
    ]
    [54,edge label={node[midway,above, sloped,font=\scriptsize]{0.58}}
     [35.1,edge label={node[midway,above, sloped,font=\scriptsize]{0.65}}
      [24.92,edge label={node[midway,above, sloped,font=\scriptsize]{0.71}}
       [23.67,edge label={node[midway,above, sloped,font=\scriptsize]{}}
        [7.34,edge label={node[midway,above, sloped,font=\scriptsize]{}}]
        [16.33,edge label={node[midway,above, sloped,font=\scriptsize]{0.69}}]
       ]
       [1.25,text=red, edge label={node[midway,above, sloped,font=\scriptsize]{0.05}}]
      ]
      [10.18,edge label={node[midway,above, sloped,font=\scriptsize]{0.29}}
       [2.95,edge label={node[midway,above, sloped,font=\scriptsize]{0.29}}
        [1.27,text=red, edge label={node[midway,above, sloped,font=\scriptsize]{}}]
        [1.68,text=red, edge label={node[midway,above, sloped,font=\scriptsize]{0.57}}]
       ]
       [7.23,edge label={node[midway,above, sloped,font=\scriptsize]{0.71}}
        [0.07,text=red, edge label={node[midway,above, sloped,font=\scriptsize]{}}]
        [7.16,edge label={node[midway,above, sloped,font=\scriptsize]{0.99}}]
       ]
      ]
     ]
     [18.9,edge label={node[midway,above, sloped,font=\scriptsize]{0.35}}
      [1.7,edge label={node[midway,above, sloped,font=\scriptsize]{0.09}}
       [0.19,text=red, edge label={node[midway,above, sloped,font=\scriptsize]{}}]
       [1.51,edge label={node[midway,above, sloped,font=\scriptsize]{0.89}}
        [1.09,text=red, edge label={node[midway,above, sloped,font=\scriptsize]{}}]
        [0.42,text=red, edge label={node[midway,above, sloped,font=\scriptsize]{0.28}}]
       ]
      ]
      [17.2,edge label={node[midway,above, sloped,font=\scriptsize]{0.91}}
       [1.2,text=red, edge label={node[midway,above, sloped,font=\scriptsize]{}}]
       [16,edge label={node[midway,above, sloped,font=\scriptsize]{0.93}}
        [4.48,edge label={node[midway,above, sloped,font=\scriptsize]{}}]
        [11.52,edge label={node[midway,above, sloped,font=\scriptsize]{0.72}}]
       ]
      ]
     ]
    ]
   ]
  ]
 ]
]    
\end{forest}
\caption{An example of the operation of the jet generator. The number on the specific node represents the total energy for a given particle, while the number on the line connecting two nodes is the energy ratio $z$ when decaying. The decay probability distribution $p(z)$ in this image is constant. The maximum number of decays in a single branch has been set to 7, and the maximum number of particles in the jet has been set to 8. A particle stops to decay once its energy is too low (here set to 0.1). The particles coloured red are removed from the jet because their energy is too low.}
\label{fig:tree}
\end{figure}

\newpage
\section{Supplementary results}
Here we give the comparison of momentum probability distributions $p^{i}_{p}(z)$ when varying the parameter $C$ as a complement to the results for $p^{i}_{E}(z)$ given in the text.
\begin{figure}[h!t!]
\vskip -4mm
\centering
\includegraphics[width=8cm]{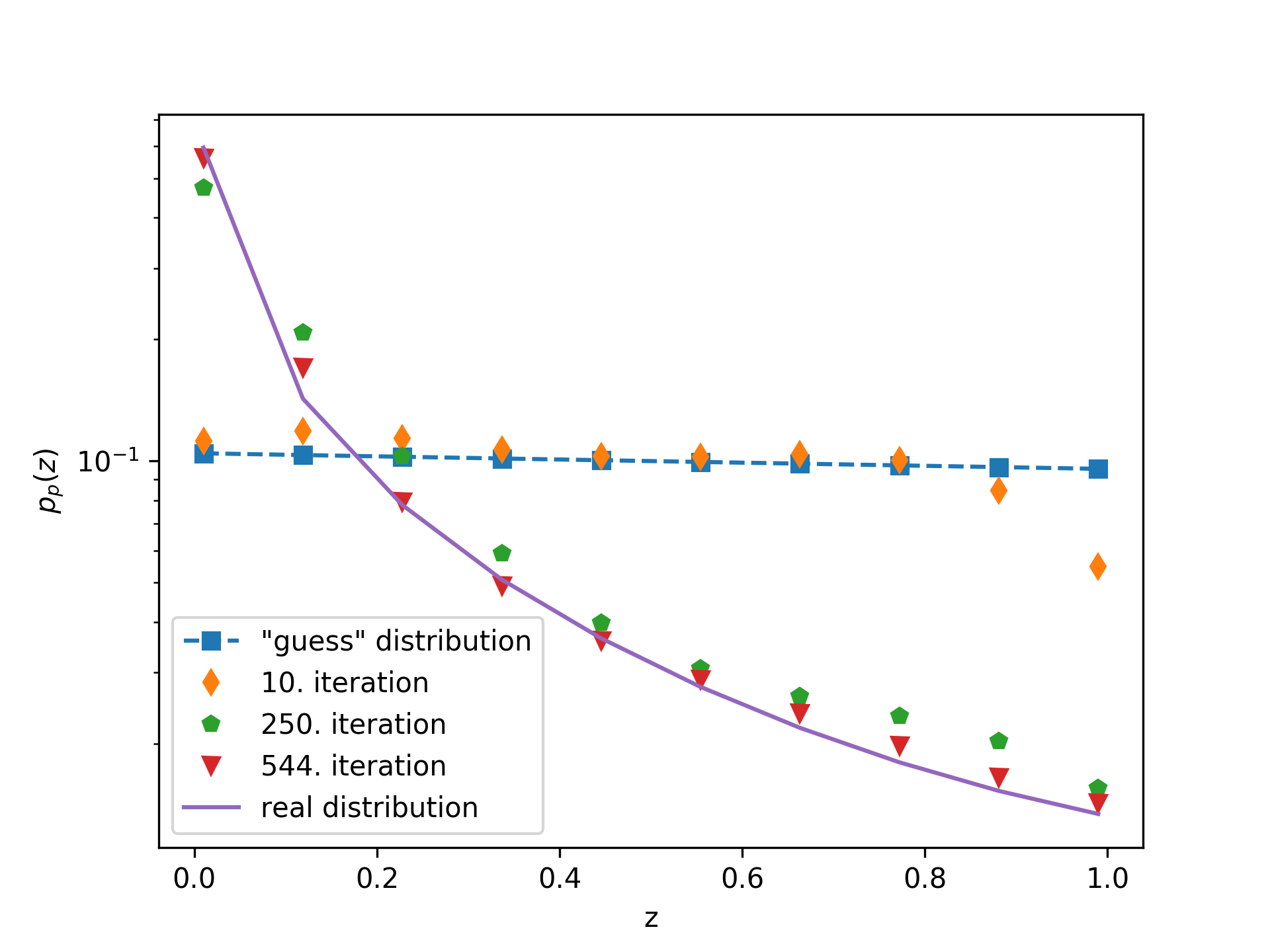}
\includegraphics[width=8cm]{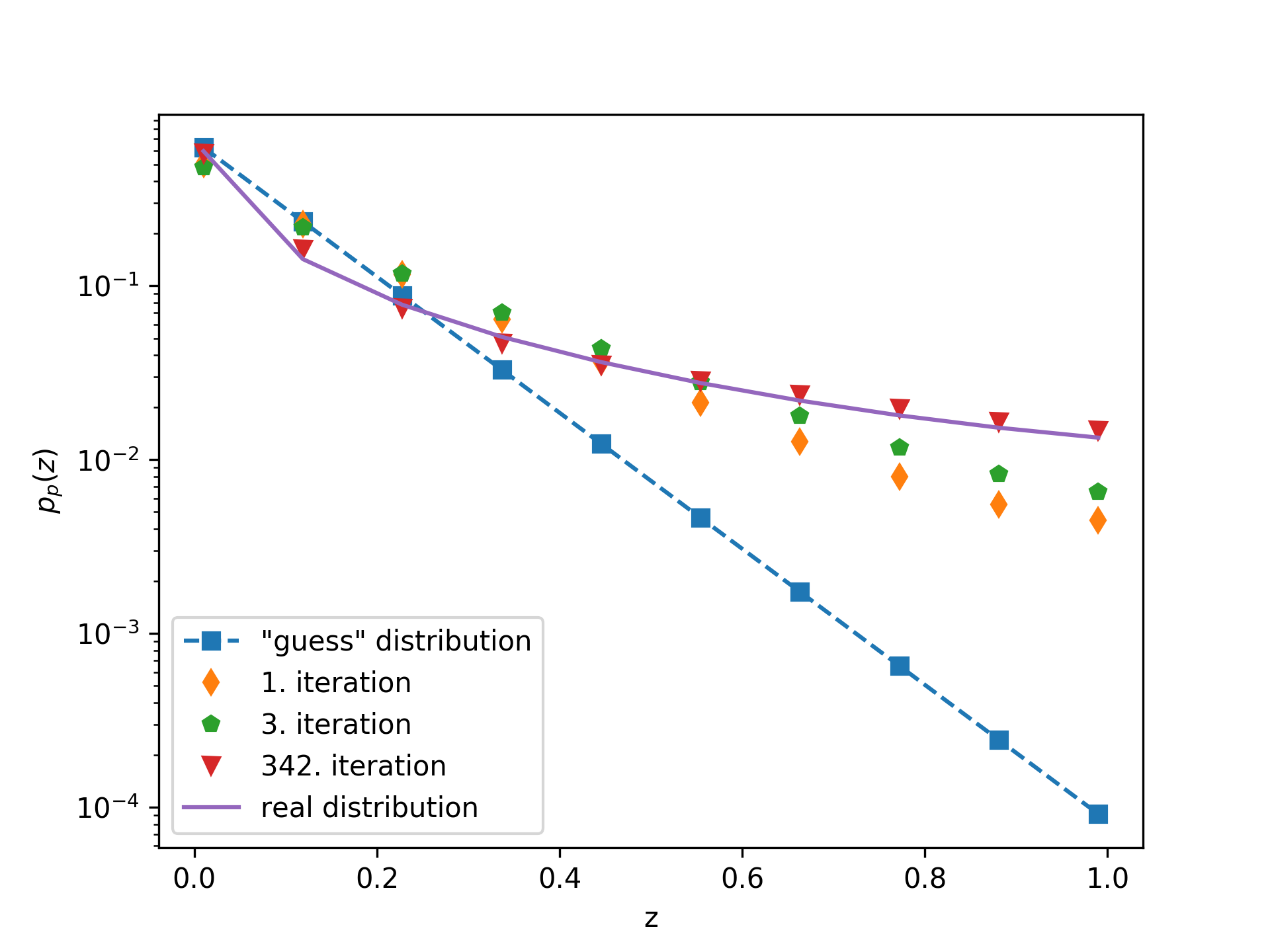}
\includegraphics[width=8cm]{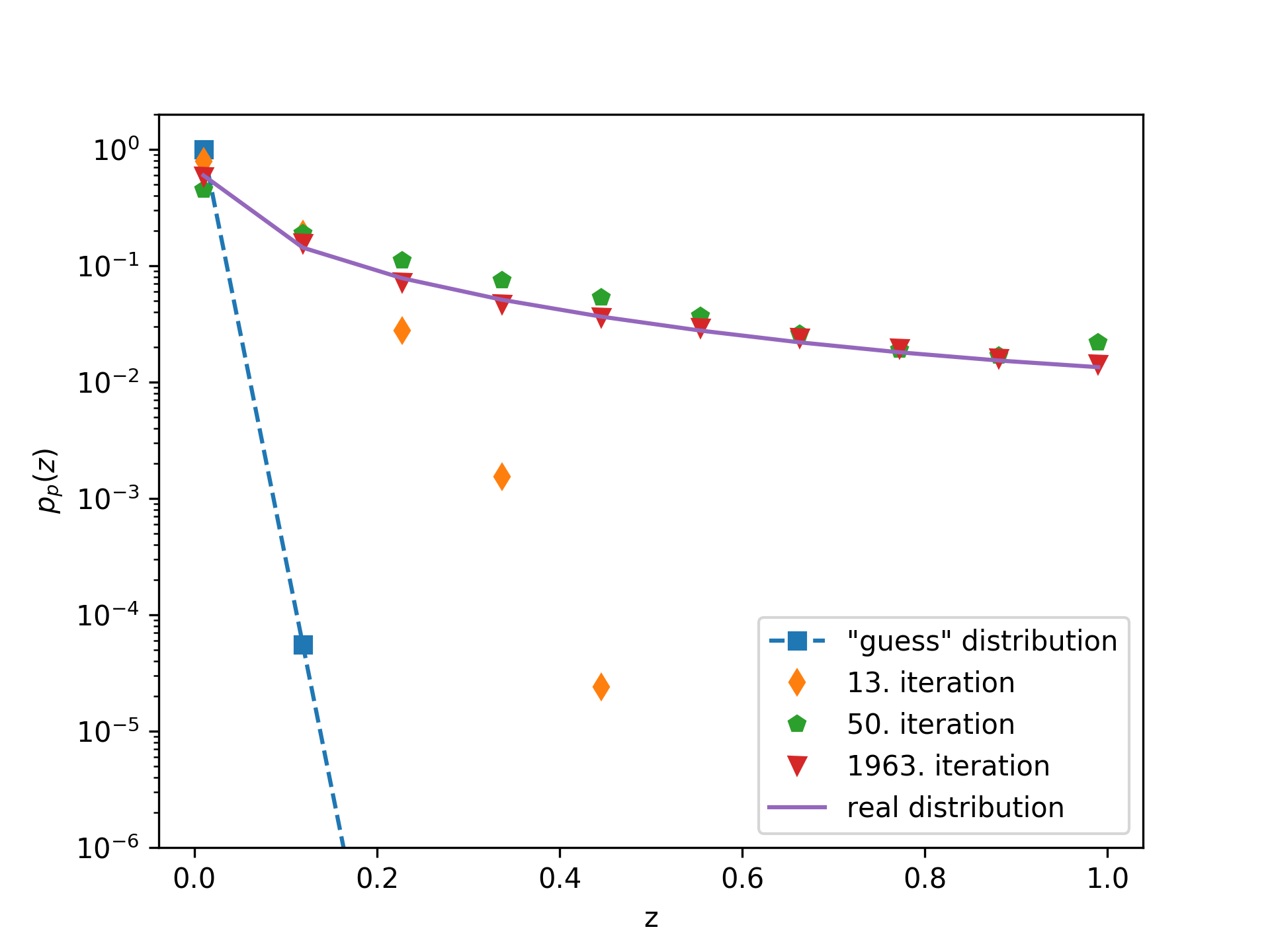}
\caption{Several iterations of the calculated probability distributions $p^{i}_{p}(z)$ (symbols) compared to $p_{\textrm{real}}(z)$ (full line) in the case of the guess distribution given by (\ref{eq_probna}). Top: $C=0.1$. Middle: $C=10$ and Bottom: $C=100$.}
\label{image_error_p}
\end{figure}   



\reftitle{References}
\newpage
\raggedright


\begin{thebibliography}{999}
\bibitem[]{pythia}
Sjostrand T, Mrenna S, Skands P. PYTHIA 6.4 Physics and Manual. arXiv:hep-ph/0603175, {\bf 2006}
\bibitem[]{NeyPear}
Neyman J, Pearson E S. On the problem of the most efficient tests of statistical hypotheses. {\em Phil. Trans. R. Soc. Lond. A.} {\bf 1933}, {\em 231}, 694--706
\bibitem[]{NNNP1}
Streit R L.A neural network for optimum Neyman-Pearson classification. IJCNN International Joint Conference on Neural Networks, {\bf 1990}, {\em 1}, 685--690
\bibitem[]{NNNP2}
Tong X, Feng Y, Li J J. Neyman-Pearson classification algorithms and NP receiver operating characteristics, {\em Sci. Adv.},{\em 4-2}, {\bf 2018}
\bibitem[]{AltPar}
Altarelli G, Parisi G. Asymptotic freedom in parton language. {\em NPB} {\bf 1977}, {\em 126}, 298--318
\bibitem[]{NNProbability}
Bishop C M. Neural networks for pattern recognition. Oxford university press, {\bf 1995}
\bibitem[]{CrossEntropy}
Nielsen M A. Neural networks and deep learning. Determination press, {\bf 2015}
\bibitem[]{RMSRE}
G\"{o}\c{c}ken M et al. Integrating metaheuristics and Artificial Neural Networks for improved stock price prediction. {\em ESA} {\bf 2016}, {\em 44}, 320--331
\bibitem[]{ErrorMargin}
Li M F et al. General models for estimating daily global solar radiation for different solar radiation zones in mainland China. {\em ECM} {\bf 2013}, {\em 70}, 139--148
\bibitem[]{geant}
Agostinelli S et al. Geant4—a simulation toolkit. {\em NIM A} {\bf 2003}, {\em 506,3}, 250--303
\bibitem[]{keras}
Chollet F et al. Keras. {\em \url{https://keras.io}}, {\bf 2015}
\bibitem[]{cnn}
LeCun Y, Bengio Y and Hinton G. Deep learning. {\em NAT} {\bf 2015}, {\em 521}, 436-444
\bibitem[]{adam}
Kingma D, Ba J. A Method for Stochastic Optimization. International Conference on Learning Representations, {\bf 2014}
\end{thebibliography}
\end{document}